\documentclass[aps,twocolumn,superscriptaddress]{revtex4}

\usepackage{graphicx}% Include figure files
\usepackage{pdfpages}
\usepackage{epstopdf}
\usepackage{amssymb}
\usepackage{mathtools}
\usepackage{color}
\usepackage{dcolumn}% Align table columns on decimal point
\usepackage{bm}% bold math
\graphicspath{{Figures/}}
\begin{document}

\title{Fluctuation-Dissipation in Thermoelectrics}

\author{N.A.M Tran}
\affiliation{Department of Electrical and Computer Engineering, McGill University, Montr\'eal, Qu\'ebec, H3A 0E9, Canada}
\author{A.S Dutt}
\affiliation{Institute for Metallic Materials, IFW Dresden, 01069 Dresden, Germany.}
\author{N.B Pulumati}
\affiliation{Institute for Metallic Materials, IFW Dresden, 01069 Dresden, Germany.}
\author{H. Reith}
\affiliation{Institute for Metallic Materials, IFW Dresden, 01069 Dresden, Germany.}
\author{A. Hu}
\affiliation{Department of Electrical and Computer Engineering, McGill University, Montr\'eal, Qu\'ebec, H3A 0E9, Canada}
\author{A. Dumont}
\affiliation{D\'epartement de physique et Institut Quantique, Universit\'e de Sherbrooke, Sherbrooke, Qu\'ebec, J1K 2R1, Canada}
\author{K. Nielsch}
\affiliation{Institute for Metallic Materials, IFW Dresden, 01069 Dresden, Germany.}
\author{A.-M. S. Tremblay}
\affiliation{D\'epartement de physique et Institut Quantique, Universit\'e de Sherbrooke, Sherbrooke, Qu\'ebec, J1K 2R1, Canada}
\author{G. Schierning}
\affiliation{Faculty of Physics, Universit\"at Bielefeld, 33501 Bielefeld, Germany.}
\author{B. Reulet}
\affiliation{D\'epartement de physique et Institut Quantique, Universit\'e de Sherbrooke, Sherbrooke, Qu\'ebec, J1K 2R1, Canada}
\author{T. Szkopek}
\affiliation{Department of Electrical and Computer Engineering, McGill University, Montr\'eal, Qu\'ebec, H3A 0E9, Canada}
\date{\today}

\begin{abstract}
Thermoelectric materials exhibit correlated transport of charge and heat. The Johnson-Nyquist noise formula $ 4 k_B T R $ for spectral density of voltage fluctuations accounts for fluctuations associated solely with Ohmic dissipation. Applying the fluctuation-dissipation theorem, we generalize the Johnson-Nyquist formula for thermoelectrics, finding an enhanced voltage fluctuation spectral density $4 k_B T R (1 + ZT)$ at frequencies below a thermal cut-off frequency $f_T$, where $ZT$ is the dimensionless thermoelectric material figure of merit. The origin of the enhancement in voltage noise is thermoelectric coupling of temperature fluctuations. We use a wideband ($f_T\sim1$~kHz), integrated thermoelectric micro-device to experimentally confirm our findings. Measuring the $ZT$ enhanced voltage noise, we experimentally resolve temperature fluctuations with an amplitude of $0.8~\mu \mathrm{K} \mathrm{Hz}^{-1/2}$ at a mean temperature of 295~K. We find that thermoelectric devices can be used for thermometry with sufficient resolution to measure the fundamental temperature fluctuations described by the fluctuation-dissipation theorem.
\end{abstract}
\maketitle

The fluctuation-dissipation theorem \cite{callen1951irreversibility} relates the fluctuations in a system at thermodynamic equilibrium with the coefficient of  irreversible dissipation under an externally applied bias field. In an electrical conductor, the fluctuation-dissipation theorem takes the form of the Johnson-Nyquist noise formula for voltage fluctuations $\langle V^2 \rangle = 4 k_B T R \Delta f$, where $k_B$ is Boltzmann's constant, $T$ is absolute temperature, $R$ is the electrical resistance, and $\Delta f$ is the electrical bandwidth \cite{johnson1928thermal,nyquist1928thermal}. The resistance $R$ that quantifies the irreversible dissipation of electrical conduction also determines the amplitude of voltage fluctuations at equilibrium. Since it's formal statement, the fluctuation-dissipation theorem has been extended to various physical systems, including the quantum optical regime \cite{gardiner2004quantum}.

Considering fluctuations in an electrical conductor further, fundamental excitations in condensed matter carry not only charge, but other physical quantities such as, for example, heat, spin, and pseudo-spin. The \textit{thermoelectric} response of a conductor is a result of the correlated transport of charge and heat by fundamental excitations. In the simplest scenario, the Seebeck coefficient $S$ is a measure of the mean entropy carried per unit charge $e$, and the Peltier coefficient $\Pi = TS$ is a measure of the mean thermal energy carried per unit charge $e$ \cite{ashcroft1976solid, rowecrc, goupil}.Correlation in the fluctuation of charge and heat transport have been considered theoretically in nanoscale systems \cite{crepieux2014}.

Surprisingly, the fluctuation-dissipation theorem as it applies to a thermoelectric material and the resulting correlations in fluctuation of charge current, heat current, voltage and temperature have not been investigated to date. Temperature fluctuation at equilibrium is well-known in the theory of statistical mechanics \cite{landau1981statistical}, but typically eludes measurement in condensed matter owing to the small absolute scale of temperature fluctuations, which scales inversely with heat capacity. Nonetheless, temperature fluctuations are of prime importance in various phenomena in condensed matter as for example they limit the sensitivity of bolometers \cite{richards1994}, generate dynamical phase transitions \cite{spahr2020} and are at the origin of non-Gaussian fluctuations in metals \cite{nagaev2002,pinsolle2018}.

In this paper, we apply the fluctuation-dissipation theorem to derive a generalized Johnson-Nyquist formula for thermoelectrics, revealing a modification in the spectral density of fluctuations that is dependent on the dimensionless, material thermoelectric figure of merit $zT=\Pi S / \kappa \rho$ (not to be confused with device $ZT$, discussed further below), where $\kappa$ is the thermal conductivity and $\rho$ is the electrical resistivity. Using integrated, wide-bandwidth, thermoelectric cooling micro-devices \cite{li2018integrated}, we measure the spectral density of voltage fluctuations and find agreement with the generalized Johnson-Nyquist formula for thermoelectrics. In the spirit of the fluctuation dissipation theorem, experimental impedance spectroscopy is further used to explicitly test the correspondence between irreversible transport coefficients and equilibrium fluctuations in a thermoelectric material.

Let us consider an electrical current $I_e$ and a heat current $I_Q$ in a thermoelectric sample of length $L$ and cross-sectional area $A$ in the presence of a time-dependent potential difference $\Delta V$ and temperature difference $\Delta T$,
\begin{equation} \label{eq:Onsager}
\begin{split}
    &I_e=L_{11}'\Delta V +L_{12}'\frac{\Delta T}{T}\\
    &I_Q=L_{21}'\Delta V +L_{22}'\frac{\Delta T}{T}+C_Q'\frac{\partial}{\partial t}\Delta T,
\end{split}
\end{equation}
where $L_{ij}'=(A/L)L_{ij}$ are the extensive transport coefficients, satisfying Onsager's reciprocal relation $L_{12}'=L_{21}'$. The intensive transport coefficients $L_{ij}$ are related to electrical conductivity $\sigma$, Seebeck coefficient $S$, thermal conductivity $\kappa$,
\begin{equation} \label{eq2}
\begin{split}
    &\sigma=L_{11}, \quad S=-\frac{1}{T}\frac{L_{12}}{L_{11}},
    \quad \kappa=\frac{1}{T}\frac{L_{22}L_{11}-L_{12}L_{21}}{L_{11}}.
\end{split}
\end{equation}
The dynamical response of the thermoelectric depends upon the extensive heat capacity $C_Q'=\partial{U}/\partial{\Delta T}$, where $U$ is the heat transported across the length $L$ of the thermoelectric.

The transport Eq.\ (1) can be compactly expressed with a generalized potential $\mathbf{\tilde{V}}$, generalized flux $\mathbf{\tilde{I}}$, and generalized conductance $\mathbb{G}$,
\begin{equation}
\begin{split}
    \mathbf{\tilde{I}}&=\mathbb{G}\mathbf{\tilde{V}},\\
    \begin{bmatrix}
        I_e(f) \\
        I_Q(f) 
    \end{bmatrix}
    &= \begin{bmatrix}
        L_{11}'&L_{12}' \\
        L_{22}'&L_{22}'+i2\pi f TC_Q'
    \end{bmatrix}
    \begin{bmatrix}
        \Delta V (f) \\
        \Delta T (f)/T 
    \end{bmatrix},
    \end{split}
    \label{Eq:OnsangerFrequency}
\end{equation}
where the Fourier transform $x(f) = \int e^{2\pi i f t} x(t) dt$ is used to work in the frequency domain.

The fluctuation-dissipation theorem applied to the generalized potential $\mathbf{\tilde{V}}$ gives,
\begin{equation}
S_{\alpha,\beta} = 4 k_B T \text{Re} \left[ \mathbb{G}_{\alpha,\beta}^{-1} \right],
  \end{equation}
where the spectral density of fluctuations $S_{\alpha,\beta}(f)$ is defined,
\begin{equation}
\begin{split}
    S_{\alpha,\beta}(f) &= \frac{2}{ \mathcal{T} } \langle \mathbf{\tilde{V}}_{\alpha} (+f) \mathbf{\tilde{V}}_{\beta} (-f) \rangle,
\end{split}
\end{equation}
where $\langle x \rangle$ denotes the ensemble average of $x$, and $\mathcal{T}=1/\Delta f$ is the reciprocal bandwidth. From an experimental perspective, $\mathcal{T}$ is the temporal duration of a measurement \cite{priestley1981spectral}, while in theoretical analysis $\mathcal{T}$ is the period in the periodic model of a stationary physical system \cite{pathria}.

%where the spectral density $S_{\alpha,\beta}(f)$ is defined in terms of the correlation function $R_{\alpha,\beta} (\tau)$,
%\begin{equation}
%\begin{split}
    %S_{\alpha,\beta}(f) &= \int_{-\infty}^{+\infty}e^{2\pi i f \tau} R_{\alpha,\beta} d\tau, \\
    %R_{\alpha,\beta}(\tau) &= \lim_{X \rightarrow \infty} \int_{-X/2}^{+X/2}\mathbf{\tilde{V}}_{\alpha} (t+\tau)\mathbf{\tilde{V}}_{\beta} (t) dt.
%\end{split}
%\end{equation}

The spectral density $S_T(f) = 2\mathcal{T}^{-1}\langle \Delta T(+f) \Delta T (-f)\rangle$ of the temperature fluctuations across the thermoelectric is,
\begin{equation}\label{Eq:TempPSD}
\begin{split}
    S_T(f) =4k_BT^3\text{Re}[\mathbb{G}_{22}^{-1}]
    =4 k_B T^2 \frac{G_T^{-1}}{1+(f/f_T)^2},
\end{split}
\end{equation}
where $G_T=\kappa A/L$ is the thermal conductance and $2\pi f_T = G_T/C'_Q$ is the thermal cut-off frequency. The total integrated temperature fluctuation, $\langle (\Delta T)^2 \rangle = \int_{0}^{\infty}S_T df =k_B T^2 / C'_Q$, in accordance with the result expected from general considerations of statistical mechanics \cite{landau1981statistical}.

The spectral density of the voltage fluctuations $S_V(f) = 2\mathcal{T}^{-1}\langle \Delta V(+f) \Delta V(-f)\rangle$ across the thermoelectric is,
\begin{equation}\label{Eq:VoltagePSD}
\begin{split}
     S_V(f)=4 k_BT\text{Re}[\mathbb{G}_{11}^{-1}] = 4  k_B T R_0+S^2 S_T(f).
    \end{split}
\end{equation}
There are voltage fluctuations of Ohmic origin proportional to the transport coefficient $R_0 = L_{11}'^{-1}$, and voltage fluctuations that originate with temperature fluctuations coupled via the thermoelectric coefficient $S$. An independent analysis showing the coupling between temperature gradient and electric field fluctuations is shown in the Supplemental Material \cite{SM}.

Combining the results of Eqs. \ref{Eq:TempPSD},\ref{Eq:VoltagePSD}, we arrive at the central theoretical result for voltage fluctuations in a thermoelectric, 
\begin{equation}\label{Eq:VoltagePSD2}
\begin{split}
    S_V(f) =4  k_B T R_0\left[ 1 + \frac{zT}{1+(f/f_T)^2} \right].
\end{split}
\end{equation}
In the high-frequency limit above the thermal cut-off frequency, $f \gg f_T$, the usual Johnson-Nyquist formula for spectral density of voltage fluctuations applies, $S_V = 4 k_BTR_0$. The spectral density of temperature fluctuations diminish $S_T(f) \propto f^{-2}$ for $f \gg f_T$. In contrast, in the low-frequency limit below thermal cut-off, $f \ll f_T$, there is an enhancement in spectral density of voltage fluctuation beyond the usual Johnson-Nyquist result, $S_V = 4 k_BTR_0 ( 1 + zT )$. The low-frequency spectral density of temperature fluctuations, $S_T \rightarrow 4 k_B T^2 G_T^{-1}$, contributes to the observable voltage fluctuations via thermoelectric coupling. The dimensionless $zT$ determines the relative enhancement of voltage noise beyond the Johnson-Nyquist result. Thermoelectric noise enhancement is thus anticipated in materials with large Seebeck coefficient $S$, large electrical conductivity $\sigma$, and small thermal conductivity $\kappa$.

\begin{figure}
    \centering
    \includegraphics[width=0.47\textwidth]{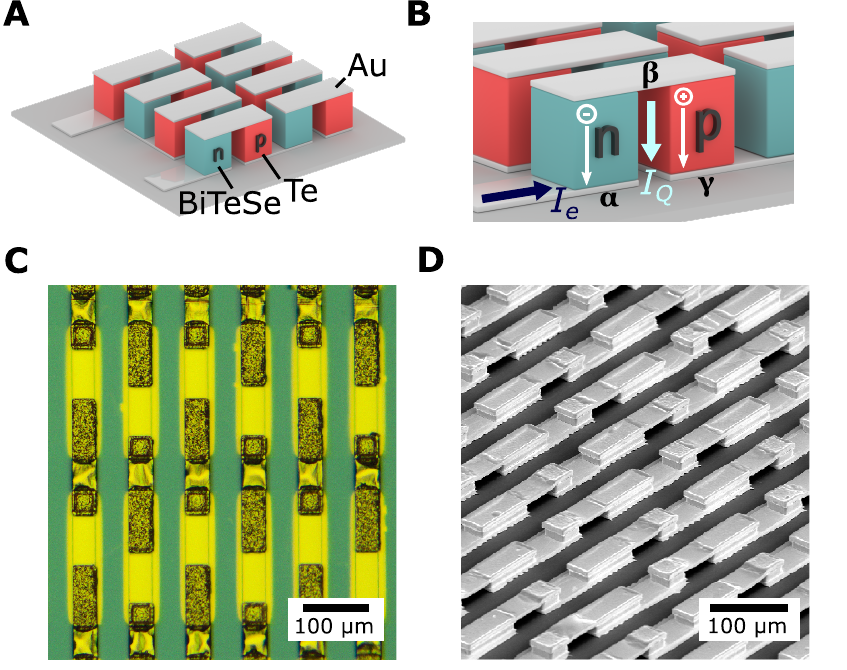}
    \caption{\textbf{Integrated thermoelectric cooler}. \textbf{A} Illustration of the alternating n-type BiTeSe and p-type Te thermoelectric legs with Au electrodes, arranged in a serpentine, series pattern on a substrate. \textbf{B} Diagram demonstrating the direction of heat current $I_Q$ downward in both n-type and p-type material as electrical current $I_e$ passes from left to right in the $\Pi$ architecture of an n/p pair.\textbf{C} Optical microscope image of the fabricated thermoelectric micro-device. \textbf{D} Oblique angle scanning electron microscope image of the fabricated thermoelectric micro-device.} 
\end{figure}

To confirm our theoretical findings, we measured the voltage noise spectral density of integrated, wide-bandwidth, thermoelectric cooling micro-devices. The devices consist of a series network of $L=10~\mu$m thick n-type (Bi$_2$(Te$_{0.95}$Se$_{0.05})_3$), abbreviated as BiTeSe) and p-type (pure Te) thermoelectric materials, with Au contacts, arranged in serpentine fashion on a Si substrate, as shown in Fig.\ 1A. A $\Pi$ architecture is adopted, Fig.\ 1B, such that the contra-oriented electrical current $I_e$ in each leg results in co-oriented heat current $I_Q$ in each leg. A microfabricated structure, Fig.\ 1C and Fig.\ 1D, was used to achieve a high thermal cut-off frequency $f_T \sim 1$~kHz, as confirmed by experiment. The cross-sectional areas of the p-type and n-type legs are $A_p=$ 85 $\mu$m $\times$ 30 $\mu$m and $A_n =$ 30 $\mu$m $\times$ 30 $\mu$m, respectively. The device systematically studied in our work consisted of 112 series leg pairs. Further details concerning the design and fabrication of wide-bandwidth thermoelectric micro-coolers have been previously reported \cite{li2018integrated}. The spectral density of voltage fluctuations for a thermolectric material in Eq.\ ( \ref{Eq:VoltagePSD2}) can be extended to a thermoelectric device consisting of a series of $N$ identical leg pairs. 

Consider first the voltage fluctuation across a single leg pair in the $\Pi$ geometry of Fig.\ 1B. Due to the effective thermal short-circuit between points $\alpha$ and $\gamma$ from the substrate thermal conductivity, the fluctuation in temperature difference $\Delta T = T_\beta -T_\alpha = T_\beta - T_\gamma$ across the n-type and p-type legs have a spectral density,
\begin{equation}
    S_{T}(f)=4k_BT^2\frac{(G_{n}+G_{p})^{-1}}{1+(f/f_{T,\Pi})^2},
\end{equation}
where $G_{n}=\kappa_n A_n/l$, $G_{p}=\kappa_p A_p/l$ are the thermal conductances of each leg, $l$ is the length of each leg, and the thermal cut-off frequency $2\pi f_{T,\Pi}=(G_{n}+G_{p})/C'_{Q,\Pi}$ where $C'_{Q,\Pi}$ is the total heat capacity for establishing a temperature difference $\Delta T$ across the leg pair. The voltage fluctuations across the n-leg ($\alpha-\beta$) and p-leg ($\beta-\gamma$) that arise from thermoelectric transduction of temperature fluctuations are correlated, and given by $(S_p-S_n)^2S_{T}(f)$. The voltage fluctuations across distinct leg pairs are not correlated, and thus the expected spectral density of voltage fluctuation across the series of $N$ leg pairs in our thermoelectric device is
\begin{equation}
    S_{V}(f)=4 k_B T N (R_n+R_p)\left[ 1+\frac{ZT}{1+(f/f_{T,\Pi})^2} \right],
    \label{Eq:VoltagePSD3}
\end{equation}
where $R_i$ and $S_i$ are the electrical resistance and Seebeck coefficient of the $i$-type leg, $i$=n,p, and $ZT$ is the thermoelectric device figure of merit,
\begin{equation}
   \label{eq:deviceFOM}
    ZT=\frac{T(S_p-S_n)^2}{(R_n+R_p) (G_n+G_p)}.
\end{equation}
The voltage fluctuation of the thermoelectric micro-device, Eq. (\ref{Eq:VoltagePSD3}), differs from the voltage fluctuation of a thermoelectric material, Eq. (\ref{Eq:VoltagePSD2}), albeit sharing a similar form. Note that the \textit{device} figure of merit $ZT$ is not the sum of the constituent \textit{material} figures of merit $zT$, due to the correlated temperature difference $\Delta T$ across p- and n-legs \cite{rowecrc}.

\begin{figure}
    \centering
    \includegraphics[width=0.47\textwidth]{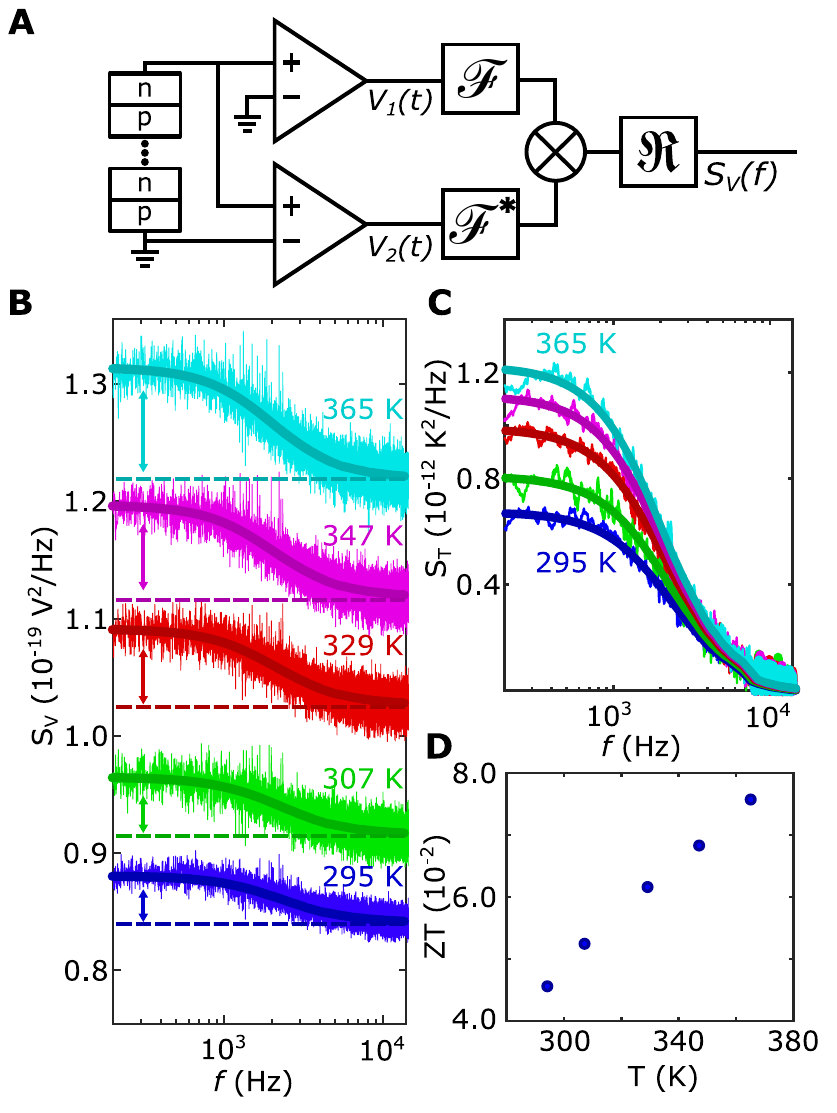}
    \caption{\textbf{Voltage and temperature fluctuations}. \textbf{A} Circuit diagram of the cross-periodogram method used to estimate the spectral density $S_V(f)$ of voltage fluctuations across the thermoelectric micro-device. Independent measurements $V_1(t)$ and $V_2(t)$ are acquired and digitally Fourier transformed to reconstruct $S_V$ from the cross-spectral density. \textbf{B} Voltage spectral density $S_V(f)$ versus frequency $f$ of the thermoelectric micro-device measured at a substrate temperature $T$ from 295~K to 365~K. Solid dark lines show a fit to theoretical spectral density of Eq.\ (\ref{Eq:VoltagePSD3}). The Ohmic Johnson-Nyquist contributions $4k_BTN(R_n+R_p)$ are indicated with dashed lines, and the excess spectral density $\Delta S_V = S_V - 4k_BTN(R_n+R_p)$ arising from thermoelectric coupling is indicated with vertical arrows. \textbf{C} The temperature fluctuation spectral density $S_T(f) = \Delta S_V/S^2$ versus frequency $f$, using the independently measured, temperature dependent, effective Seebeck coefficient $S$. A 40 Hz running average was applied to $S_T$ for visual clarity, and a model fit to Eq.\ (9) is shown. \textbf{D} The thermoelectric micro-device figure of merit $ZT=\lim_{f \rightarrow 0}\Delta S_V/(S_V-\Delta S_V)$ versus substrate temperature $T$.}
\end{figure}

The voltage power spectral density $S_V(f)$ was measured using a correlation method, as shown in Fig.\ 2A. The voltage across two terminals of a series of p/n leg pairs was measured simultaneously with two cascaded voltage amplifiers. Each cascade consisted of two identical voltage pre-amplifiers (LI-75, NF Corp.) with 40 dB total gain. The voltages $V_1(t)$ and $V_2(t)$ were digitized (DT-9847, Data Translation) and the cross-spectral density estimated using the cross-periodogram method \cite{priestley1981spectral}. To suppress the uncorrelated amplifier noise in $V_1(t)$ and $V_2(t)$, we averaged $7.5\times10^4$ periodograms, with $2.15\times10^5$ samples/periodogram acquired at a sampling rate of $2.15\times10^5$ samples/s, for over 20 hours of acquisition time per spectral density measurement. The micro-device was placed in a series of metal boxes for electrical and thermal isolation, open-loop resistive heating was applied to vary the sample substrate temperature $T$, and a thermistor was used to measure the $T$, which varied by less than $\pm1$~K over the duration of each measurement.

The spectral density of the voltage across the thermoelectric is reconstructed as $S_V = 2\mathcal{T}^{-1}\text{Re} \left\{ V_1(f)V^*_2(f)\right\}$. The comparatively high thermal cut-off frequency, $f_{T,\Pi}\sim1$~kHz, is essential to avoid measurements in the extremely low frequency (ELF, $f<30$~Hz) band where amplifier $1/f$ noise is prohibitively large. Frequency independent measurement system response over the bandwidth $200~\mathrm{Hz}<f<20~\mathrm{kHz}$ was confirmed using a resistor of $R = 25.9~\Omega$ (see Supplemental Material). The gain in the noise measurement system was calibrated with an $R=8.20~\Omega$ resistor at $f=1.7$~kHz.

The measured voltage spectral density $S_V(f)$ versus frequency $f$ is shown in Fig.\ 2B, along with a numerical fit to Eq.\ (\ref{Eq:VoltagePSD3}). Measurements were taken with the thermoelectric substrate temperature $T$ varied from 295 K to 365 K, adjusted with an external resistive heater and measured via thermistor. The dashed horizontal lines denote the Ohmic Johnson-Nyquist noise contribution, $4k_BTN(R_n+R_p)$, with the resistance $N(R_n+R_p)$ determined from the fit of the measured $S_V$ to Eq.\ (\ref{Eq:VoltagePSD3}). There is an enhancement in voltage fluctuations $\Delta S_V=S_V - 4k_BTN(R_n+R_p)$ in the low-frequency limit $f \ll f_T$ as compared to the high-frequency limit $f \gg f_T$, indicated with vertical arrows.

\begin{figure}
    \centering
    \includegraphics[width=0.47\textwidth]{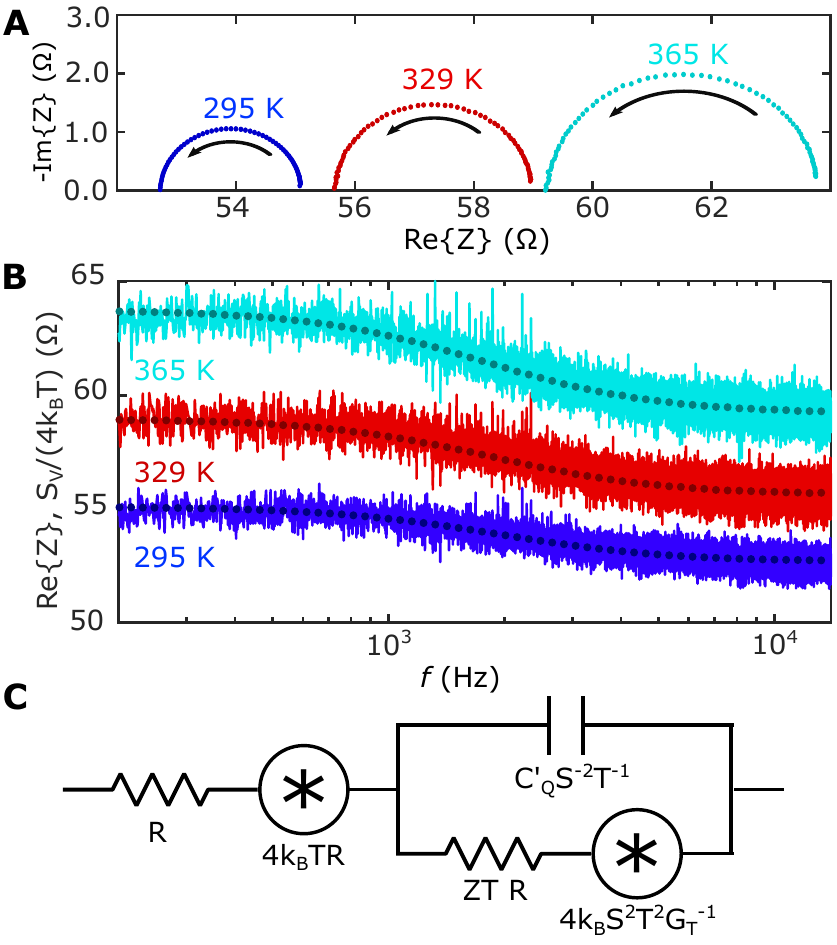}
    \caption{\textbf{Impedance and noise spectroscopy}. \textbf{A} Nyquist plot of the measured impedance, $-\mathrm{Im}\left\{Z\right\}$ versus $\mathrm{Re}\left\{Z\right\}$ at $T=$295~K,329~K, and 365~K, over the frequency $100~\mathrm{Hz}<f<100~\mathrm{kHz}$. The semi-circle arrows direction indicates an increase in frequency. \textbf{B} A comparison of the measured dissipative impedance $\mathrm{Re}\left\{Z\right\}$ (darker circles) and the measured voltage spectral density normalized to impedance units, $S_V / 4 k_B T$, (thin line) versus frequency $f$. \textbf{C} Model circuit for a thermoelectric, including Ohmic and thermoelectric dissipative elements, Ohmic and thermolectric fluctuation sources, and effective capacitance. }
    \label{fig:TElegpair}
\end{figure}

The spectral density of temperature fluctuations, $S_T=\Delta S_V/S^2$, was determined using the effective, temperature dependent, Seebeck coefficient $S = S_p-S_n$ inferred from measurements of BiTeSe and Te thin films (see Supplemental Material). For example, at $T=298~$K, $S_p=194~\mu\mathrm{V}\mathrm{K}^{-1}$, $S_n = -44~\mu\mathrm{V}\mathrm{K}^{-1}$ and $S=~238~\mu\mathrm{V}\mathrm{K}^{-1}$. The experimentally determined spectral density of temperature fluctuations, $S_T$, is shown in Fig.\ 2C versus substrate temperature $T$, along with a model fit to Eq.\ (9). At $T=295~$K, the experimentally resolved amplitude of temperature fluctuations is $\delta T = S_T^{1/2}=0.8~\mu \mathrm{K}\mathrm{Hz}^{-1/2}$, corresponding to a relative fluctuation amplitude $\delta T/T = 3\times10^{-9}~\mathrm{Hz}^{-1/2}$. In the context of thermometry, our findings show that a thermoelectric micro-device can be used to measure temperature differences as small as the fundamental fluctuations described by the fluctuation-dissipation theorem.

The micro-device figure of merit $ZT$ can be determined from the frequency dependent voltage noise $S_V$, with $ZT = \lim_{f\rightarrow 0} \Delta S_V/(S_V-\Delta S_V)$. The figure of merit $ZT$ versus substrate temperature $T$ is shown in Fig.\ 2D, where $ZT$ was determined from a fit of the experimentally measured $S_V$ to Eq.\ (10). As anticipated, $ZT$ increases with substrate temperature $T$.

The fluctuation dissipation theorem identifies the equivalency of the coefficient for fluctuation at thermal equilibrium and the coefficient for dissipative transport. We therefore investigated this equivalency for thermoelectrics. The transport Eq.\ (1) leads to a frequency dependent electrical impedance,
\begin{equation}
    Z(f)=\frac{\Delta V}{I_e}=R \left( 1+\frac{ZT}{1+i f/f_{T,\Pi}} \right),
\label{Eq:Z}
\end{equation}
where $R=N(R_n+R_p)$ for our thermoelectric micro-device. The impedance of Eq.\ (\ref{Eq:Z}) is the basis for impedance spectroscopy of thermoelectrics, which have evolved in complexity \cite{harman1958special,downey2007characterization, garcia2014impedance,garcia2016thermal,hasegawa2016thermoelectric}. The thermoelectric cooling device impedance was measured using a lock-in amplifier (MLFI 500kHz, Zurich Instruments) with an AC current bias $I_e$ of 100 $\mu$A amplitude over a frequency range $100~\mathrm{Hz}<f<100~\mathrm{kHz}$. The Nyquist plot of measured $-\mathrm{Im}\left\{Z\right\}$ versus $\mathrm{Re}\left\{Z\right\}$ is shown in Fig.\ 3A at $T=$295~K, 329~K, and 365~K. The semi-circle characteristic of first-order response is evident in the Nyquist plot, and a numerical fit to the simple model of Eq.\ (\ref{Eq:Z}). A direct comparison of the measured dissipative impedance coefficient $\mathrm{Re}\left\{Z\right\}$ and the measured normalized voltage fluctuation spectral density $S_V / 4 k_B T$ is shown in Fig.\ 3B. Our experiments are in good agreement with the fluctuation dissipation theorem as applied to the effective electronic sector of the transport equations, $\mathrm{Re}\left\{Z\right\}=S_V / 4 k_B T$. The thermoelectric coupling thus simultaneously enhances both the spectral density of voltage fluctuations and the electrical impedance above that of the purely Ohmic contribution to each. Voltage fluctuations are enhanced by thermoelectric coupling of temperature fluctuations. Impedance is enhanced by the voltage generated by thermoelectric coupling to the temperature gradient established by the passage of electrical (and thus heat) current in the thermoelectric.

Assembling our findings, a simple equivalent electrical transport model for a thermoelectric element is shown in Fig.\ 3C, using simplified notation. The model includes the Ohmic resistance $R$ and associated Johnson-Nyquist fluctuations of spectral density $4 k_B TR$, a resistance $ZT \cdot R = T S^2 G_T^{-1}$ of thermoelectric origin with an associated spectral density of voltage fluctuations $4k_B T R \cdot ZT = 4k_B S^2 T^2 G_T^{-1}$. The frequency dependence of thermal contributions to impedance and voltage fluctuations is accounted for with the inclusion of an effective capacitance $C_T=C'_Q T^{-1}S^{-2}$ of thermal origin, with cut-off frequency $2 \pi f_{T} = G_T/C_T$.

In conclusion, we have demonstrated how the fluctuation dissipation theorem can be extended to thermoelectrics, revealing the role of temperature fluctuations in the modification of the renowned Johnson-Nyquist formula. Our work establishes a quantitative understanding of temperature fluctuations in thermoelectrics, defining an important physical limit for micro-scale thermoelectric thermometery in, for example, intracellular thermometry \cite{tian2015high}. Finally, we note that our results can be generalized to include other physical quantities carried by fundamental excitations in a conductor. For example, spin-orbit coupling can result in fundamental excitations that carry charge and spin, and the correlated transport of charge and spin can in principle lead to a modification of the observable spectral density of voltage fluctuations. Moreover, the spectral density of voltage fluctuations may give insight into the spectral density of spin fluctuations.

\textbf{Acknowledgements}
A.-M.S.T., B.R. and T.S. acknowledge financial support from the Natural Sciences and Engineering Research Council of Canada (NSERC) under grants RGPIN-2016-04400, RGPIN-2018-04851, RGPIN-2019-05312, the Canada First Research Excellence Fund, and the Canada Excellence Research Chairs program. A.S.D., H.R. and G.S. acknowledge funding from the DFG (Deutsche Forschungsgemeinschaft) under grant number SCHI1010/11-1, and N.B.P. and H.R. acknowledge funding under grant number RE3973/1-1.

\bibliographystyle{apsrev4-2}
\bibliography{references}

%apsrev4-2.bst 2019-01-14 (MD) hand-edited version of apsrev4-1.bst
%Control: key (0)
%Control: author (72) initials jnrlst
%Control: editor formatted (1) identically to author
%Control: production of article title (-1) disabled
%Control: page (0) single
%Control: year (1) truncated
%Control: production of eprint (0) enabled
\begin{thebibliography}{23}%
\makeatletter
\providecommand \@ifxundefined [1]{%
 \@ifx{#1\undefined}
}%
\providecommand \@ifnum [1]{%
 \ifnum #1\expandafter \@firstoftwo
 \else \expandafter \@secondoftwo
 \fi
}%
\providecommand \@ifx [1]{%
 \ifx #1\expandafter \@firstoftwo
 \else \expandafter \@secondoftwo
 \fi
}%
\providecommand \natexlab [1]{#1}%
\providecommand \enquote  [1]{``#1''}%
\providecommand \bibnamefont  [1]{#1}%
\providecommand \bibfnamefont [1]{#1}%
\providecommand \citenamefont [1]{#1}%
\providecommand \href@noop [0]{\@secondoftwo}%
\providecommand \href [0]{\begingroup \@sanitize@url \@href}%
\providecommand \@href[1]{\@@startlink{#1}\@@href}%
\providecommand \@@href[1]{\endgroup#1\@@endlink}%
\providecommand \@sanitize@url [0]{\catcode `\\12\catcode `\$12\catcode
  `\&12\catcode `\#12\catcode `\^12\catcode `\_12\catcode `\%12\relax}%
\providecommand \@@startlink[1]{}%
\providecommand \@@endlink[0]{}%
\providecommand \url  [0]{\begingroup\@sanitize@url \@url }%
\providecommand \@url [1]{\endgroup\@href {#1}{\urlprefix }}%
\providecommand \urlprefix  [0]{URL }%
\providecommand \Eprint [0]{\href }%
\providecommand \doibase [0]{https://doi.org/}%
\providecommand \selectlanguage [0]{\@gobble}%
\providecommand \bibinfo  [0]{\@secondoftwo}%
\providecommand \bibfield  [0]{\@secondoftwo}%
\providecommand \translation [1]{[#1]}%
\providecommand \BibitemOpen [0]{}%
\providecommand \bibitemStop [0]{}%
\providecommand \bibitemNoStop [0]{.\EOS\space}%
\providecommand \EOS [0]{\spacefactor3000\relax}%
\providecommand \BibitemShut  [1]{\csname bibitem#1\endcsname}%
\let\auto@bib@innerbib\@empty
%</preamble>
\bibitem [{\citenamefont {Callen}\ and\ \citenamefont
  {Welton}(1951)}]{callen1951irreversibility}%
  \BibitemOpen
  \bibfield  {author} {\bibinfo {author} {\bibfnamefont {H.~B.}\ \bibnamefont
  {Callen}}\ and\ \bibinfo {author} {\bibfnamefont {T.~A.}\ \bibnamefont
  {Welton}},\ }\href@noop {} {\bibfield  {journal} {\bibinfo  {journal}
  {Physical Review}\ }\textbf {\bibinfo {volume} {83}},\ \bibinfo {pages} {34}
  (\bibinfo {year} {1951})}\BibitemShut {NoStop}%
\bibitem [{\citenamefont {Johnson}(1928)}]{johnson1928thermal}%
  \BibitemOpen
  \bibfield  {author} {\bibinfo {author} {\bibfnamefont {J.~B.}\ \bibnamefont
  {Johnson}},\ }\href@noop {} {\bibfield  {journal} {\bibinfo  {journal}
  {Physical review}\ }\textbf {\bibinfo {volume} {32}},\ \bibinfo {pages} {97}
  (\bibinfo {year} {1928})}\BibitemShut {NoStop}%
\bibitem [{\citenamefont {Nyquist}(1928)}]{nyquist1928thermal}%
  \BibitemOpen
  \bibfield  {author} {\bibinfo {author} {\bibfnamefont {H.}~\bibnamefont
  {Nyquist}},\ }\href@noop {} {\bibfield  {journal} {\bibinfo  {journal}
  {Physical review}\ }\textbf {\bibinfo {volume} {32}},\ \bibinfo {pages} {110}
  (\bibinfo {year} {1928})}\BibitemShut {NoStop}%
\bibitem [{\citenamefont {Gardiner}\ and\ \citenamefont
  {Zoller}(2004)}]{gardiner2004quantum}%
  \BibitemOpen
  \bibfield  {author} {\bibinfo {author} {\bibfnamefont {C.}~\bibnamefont
  {Gardiner}}\ and\ \bibinfo {author} {\bibfnamefont {P.}~\bibnamefont
  {Zoller}},\ }\href@noop {} {\emph {\bibinfo {title} {Quantum noise}}}\
  (\bibinfo  {publisher} {Springer, New York},\ \bibinfo {year}
  {2004})\BibitemShut {NoStop}%
\bibitem [{\citenamefont {Ashcroft}\ \emph {et~al.}(1976)\citenamefont
  {Ashcroft}, \citenamefont {Mermin} \emph {et~al.}}]{ashcroft1976solid}%
  \BibitemOpen
  \bibfield  {author} {\bibinfo {author} {\bibfnamefont {N.~W.}\ \bibnamefont
  {Ashcroft}}, \bibinfo {author} {\bibfnamefont {N.~D.}\ \bibnamefont
  {Mermin}}, \emph {et~al.},\ }\href@noop {} {\emph {\bibinfo {title} {Solid
  state physics}}},\ Vol.\ \bibinfo {volume} {2005}\ (\bibinfo  {publisher}
  {Holt, Rinehart and Winston, New York},\ \bibinfo {year} {1976})\BibitemShut
  {NoStop}%
\bibitem [{\citenamefont {Rowe}(1995)}]{rowecrc}%
  \BibitemOpen
  \bibfield  {author} {\bibinfo {author} {\bibfnamefont {D.~M.}\ \bibnamefont
  {Rowe}},\ }\href@noop {} {\emph {\bibinfo {title} {CRC Handbook of
  Thermoelectrics}}}\ (\bibinfo  {publisher} {CRC Press, Boca Raton},\ \bibinfo
  {year} {1995})\BibitemShut {NoStop}%
\bibitem [{\citenamefont {Goupil}\ \emph {et~al.}(2011)\citenamefont {Goupil},
  \citenamefont {Seifert}, \citenamefont {Zabrocki}, \citenamefont
  {M\"{u}ller},\ and\ \citenamefont {Snyder}}]{goupil}%
  \BibitemOpen
  \bibfield  {author} {\bibinfo {author} {\bibfnamefont {C.}~\bibnamefont
  {Goupil}}, \bibinfo {author} {\bibfnamefont {W.}~\bibnamefont {Seifert}},
  \bibinfo {author} {\bibfnamefont {K.}~\bibnamefont {Zabrocki}}, \bibinfo
  {author} {\bibfnamefont {E.}~\bibnamefont {M\"{u}ller}},\ and\ \bibinfo
  {author} {\bibfnamefont {G.~J.}\ \bibnamefont {Snyder}},\ }{\bibfield  {journal} {\bibinfo
  {journal} {Entropy}\ }\textbf {\bibinfo {volume} {13}},\ \bibinfo {pages}
  {1481} (\bibinfo {year} {2011})}\BibitemShut {NoStop}%
\bibitem [{\citenamefont {Cr{\'e}pieux}\ and\ \citenamefont
  {Michelini}(2014)}]{crepieux2014}%
  \BibitemOpen
  \bibfield  {author} {\bibinfo {author} {\bibfnamefont {A.}~\bibnamefont
  {Cr{\'e}pieux}}\ and\ \bibinfo {author} {\bibfnamefont {F.}~\bibnamefont
  {Michelini}},\ }\href@noop {} {\bibfield  {journal} {\bibinfo  {journal}
  {Journal of Physics: Condensed Matter}\ }\textbf {\bibinfo {volume} {27}},\
  \bibinfo {pages} {015302} (\bibinfo {year} {2014})}\BibitemShut {NoStop}%
\bibitem [{\citenamefont {Landau}\ and\ \citenamefont
  {Lifshitz}(1981)}]{landau1981statistical}%
  \BibitemOpen
  \bibfield  {author} {\bibinfo {author} {\bibfnamefont {L.~D.}\ \bibnamefont
  {Landau}}\ and\ \bibinfo {author} {\bibfnamefont {E.~M.}\ \bibnamefont
  {Lifshitz}},\ }\href@noop {} {\emph {\bibinfo {title} {Course of Theoretical
  Physics: Statistical Physics Part 1}}}\ (\bibinfo  {publisher} {Elsevier},\
  \bibinfo {year} {1981})\BibitemShut {NoStop}%
\bibitem [{\citenamefont {Richards}(1994)}]{richards1994}%
  \BibitemOpen
  \bibfield  {author} {\bibinfo {author} {\bibfnamefont {P.}~\bibnamefont
  {Richards}},\ }\href@noop {} {\bibfield  {journal} {\bibinfo  {journal}
  {Journal of Applied Physics}\ }\textbf {\bibinfo {volume} {76}},\ \bibinfo
  {pages} {1} (\bibinfo {year} {1994})}\BibitemShut {NoStop}%
\bibitem [{\citenamefont {Spahr}\ \emph {et~al.}(2020)\citenamefont {Spahr},
  \citenamefont {Graveline}, \citenamefont {Lupien}, \citenamefont {Aprili},\
  and\ \citenamefont {Reulet}}]{spahr2020}%
  \BibitemOpen
  \bibfield  {author} {\bibinfo {author} {\bibfnamefont {K.}~\bibnamefont
  {Spahr}}, \bibinfo {author} {\bibfnamefont {J.}~\bibnamefont {Graveline}},
  \bibinfo {author} {\bibfnamefont {C.}~\bibnamefont {Lupien}}, \bibinfo
  {author} {\bibfnamefont {M.}~\bibnamefont {Aprili}},\ and\ \bibinfo {author}
  {\bibfnamefont {B.}~\bibnamefont {Reulet}},\ }\href@noop {} {\bibfield
  {journal} {\bibinfo  {journal} {Physical Review B}\ }\textbf {\bibinfo
  {volume} {102}},\ \bibinfo {pages} {100504} (\bibinfo {year}
  {2020})}\BibitemShut {NoStop}%
\bibitem [{\citenamefont {Nagaev}(2002)}]{nagaev2002}%
  \BibitemOpen
  \bibfield  {author} {\bibinfo {author} {\bibfnamefont {K.}~\bibnamefont
  {Nagaev}},\ }\href@noop {} {\bibfield  {journal} {\bibinfo  {journal}
  {Physical Review B}\ }\textbf {\bibinfo {volume} {66}},\ \bibinfo {pages}
  {075334} (\bibinfo {year} {2002})}\BibitemShut {NoStop}%
\bibitem [{\citenamefont {Pinsolle}\ \emph {et~al.}(2018)\citenamefont
  {Pinsolle}, \citenamefont {Houle}, \citenamefont {Lupien},\ and\
  \citenamefont {Reulet}}]{pinsolle2018}%
  \BibitemOpen
  \bibfield  {author} {\bibinfo {author} {\bibfnamefont {E.}~\bibnamefont
  {Pinsolle}}, \bibinfo {author} {\bibfnamefont {S.}~\bibnamefont {Houle}},
  \bibinfo {author} {\bibfnamefont {C.}~\bibnamefont {Lupien}},\ and\ \bibinfo
  {author} {\bibfnamefont {B.}~\bibnamefont {Reulet}},\ }\href@noop {}
  {\bibfield  {journal} {\bibinfo  {journal} {Physical review letters}\
  }\textbf {\bibinfo {volume} {121}},\ \bibinfo {pages} {027702} (\bibinfo
  {year} {2018})}\BibitemShut {NoStop}%
\bibitem [{\citenamefont {Li}\ \emph {et~al.}(2018)\citenamefont {Li},
  \citenamefont {Fernandez}, \citenamefont {Ramos}, \citenamefont {Barati},
  \citenamefont {P{\'e}rez}, \citenamefont {Soldatov}, \citenamefont {Reith},
  \citenamefont {Schierning},\ and\ \citenamefont
  {Nielsch}}]{li2018integrated}%
  \BibitemOpen
  \bibfield  {author} {\bibinfo {author} {\bibfnamefont {G.}~\bibnamefont
  {Li}}, \bibinfo {author} {\bibfnamefont {J.~G.}\ \bibnamefont {Fernandez}},
  \bibinfo {author} {\bibfnamefont {D.~A.~L.}\ \bibnamefont {Ramos}}, \bibinfo
  {author} {\bibfnamefont {V.}~\bibnamefont {Barati}}, \bibinfo {author}
  {\bibfnamefont {N.}~\bibnamefont {P{\'e}rez}}, \bibinfo {author}
  {\bibfnamefont {I.}~\bibnamefont {Soldatov}}, \bibinfo {author}
  {\bibfnamefont {H.}~\bibnamefont {Reith}}, \bibinfo {author} {\bibfnamefont
  {G.}~\bibnamefont {Schierning}},\ and\ \bibinfo {author} {\bibfnamefont
  {K.}~\bibnamefont {Nielsch}},\ }\href@noop {} {\bibfield  {journal} {\bibinfo
   {journal} {Nature Electronics}\ }\textbf {\bibinfo {volume} {1}},\ \bibinfo
  {pages} {555} (\bibinfo {year} {2018})}\BibitemShut {NoStop}%
\bibitem [{\citenamefont {Priestley}(1981)}]{priestley1981spectral}%
  \BibitemOpen
  \bibfield  {author} {\bibinfo {author} {\bibfnamefont {M.~B.}\ \bibnamefont
  {Priestley}},\ }\href@noop {} {\emph {\bibinfo {title} {Spectral analysis and
  time series: probability and mathematical statistics}}}\ (\bibinfo
  {publisher} {Academic Press, San Diego},\ \bibinfo {year} {1981})\BibitemShut
  {NoStop}%
\bibitem [{\citenamefont {Pathria}(1996)}]{pathria}%
  \BibitemOpen
  \bibfield  {author} {\bibinfo {author} {\bibfnamefont {R.~K.}\ \bibnamefont
  {Pathria}},\ }\href@noop {} {\emph {\bibinfo {title} {Statistical
  Mechanics}}}\ (\bibinfo  {publisher} {Butterworth-Heineman, Oxford},\
  \bibinfo {year} {1996})\BibitemShut {NoStop}%
\bibitem [{SM()}]{SM}%
  \BibitemOpen
  \href@noop {} {\bibinfo {title} {Supplemental material is available,
  including theoretical analysis of electric field and temperature gradient
  fluctuations, voltage fluctuation measurement of a thin film resistor,
  voltage fluctuation measurement of a {N}=105 thermoelectric micro-device,
  numerical model fits of voltage fluctuation and impedance spectra, {S}eebeck
  coefficient measurements.}}\BibitemShut {Stop}%
\bibitem [{\citenamefont {Harman}(1958)}]{harman1958special}%
  \BibitemOpen
  \bibfield  {author} {\bibinfo {author} {\bibfnamefont {T.~C.}\ \bibnamefont
  {Harman}},\ }\href@noop {} {\bibfield  {journal} {\bibinfo  {journal}
  {Journal of Applied Physics}\ }\textbf {\bibinfo {volume} {29}},\ \bibinfo
  {pages} {1373} (\bibinfo {year} {1958})}\BibitemShut {NoStop}%
\bibitem [{\citenamefont {Downey}\ \emph {et~al.}(2007)\citenamefont {Downey},
  \citenamefont {Hogan},\ and\ \citenamefont
  {Cook}}]{downey2007characterization}%
  \BibitemOpen
  \bibfield  {author} {\bibinfo {author} {\bibfnamefont {A.~D.}\ \bibnamefont
  {Downey}}, \bibinfo {author} {\bibfnamefont {T.~P.}\ \bibnamefont {Hogan}},\
  and\ \bibinfo {author} {\bibfnamefont {B.}~\bibnamefont {Cook}},\ }\href@noop
  {} {\bibfield  {journal} {\bibinfo  {journal} {Review of Scientific
  Instruments}\ }\textbf {\bibinfo {volume} {78}},\ \bibinfo {pages} {093904}
  (\bibinfo {year} {2007})}\BibitemShut {NoStop}%
\bibitem [{\citenamefont {Garc{\'\i}a-Ca{\~n}adas}\ and\ \citenamefont
  {Min}(2014)}]{garcia2014impedance}%
  \BibitemOpen
  \bibfield  {author} {\bibinfo {author} {\bibfnamefont {J.}~\bibnamefont
  {Garc{\'\i}a-Ca{\~n}adas}}\ and\ \bibinfo {author} {\bibfnamefont
  {G.}~\bibnamefont {Min}},\ }\href@noop {} {\bibfield  {journal} {\bibinfo
  {journal} {Journal of Applied Physics}\ }\textbf {\bibinfo {volume} {116}},\
  \bibinfo {pages} {174510} (\bibinfo {year} {2014})}\BibitemShut {NoStop}%
\bibitem [{\citenamefont {Garc{\'\i}a-Ca{\~n}adas}\ and\ \citenamefont
  {Min}(2016)}]{garcia2016thermal}%
  \BibitemOpen
  \bibfield  {author} {\bibinfo {author} {\bibfnamefont {J.}~\bibnamefont
  {Garc{\'\i}a-Ca{\~n}adas}}\ and\ \bibinfo {author} {\bibfnamefont
  {G.}~\bibnamefont {Min}},\ }\href@noop {} {\bibfield  {journal} {\bibinfo
  {journal} {AIP Advances}\ }\textbf {\bibinfo {volume} {6}},\ \bibinfo {pages}
  {035008} (\bibinfo {year} {2016})}\BibitemShut {NoStop}%
\bibitem [{\citenamefont {Hasegawa}\ \emph {et~al.}(2016)\citenamefont
  {Hasegawa}, \citenamefont {Homma},\ and\ \citenamefont
  {Ohtsuka}}]{hasegawa2016thermoelectric}%
  \BibitemOpen
  \bibfield  {author} {\bibinfo {author} {\bibfnamefont {Y.}~\bibnamefont
  {Hasegawa}}, \bibinfo {author} {\bibfnamefont {R.}~\bibnamefont {Homma}},\
  and\ \bibinfo {author} {\bibfnamefont {M.}~\bibnamefont {Ohtsuka}},\
  }\href@noop {} {\bibfield  {journal} {\bibinfo  {journal} {Journal of
  Electronic Materials}\ }\textbf {\bibinfo {volume} {45}},\ \bibinfo {pages}
  {1886} (\bibinfo {year} {2016})}\BibitemShut {NoStop}%
\bibitem [{\citenamefont {Tian}\ \emph {et~al.}(2015)\citenamefont {Tian},
  \citenamefont {Wang}, \citenamefont {Wang}, \citenamefont {Chen},
  \citenamefont {Sun}, \citenamefont {Li}, \citenamefont {Wang},\ and\
  \citenamefont {Gu}}]{tian2015high}%
  \BibitemOpen
  \bibfield  {author} {\bibinfo {author} {\bibfnamefont {W.}~\bibnamefont
  {Tian}}, \bibinfo {author} {\bibfnamefont {C.}~\bibnamefont {Wang}}, \bibinfo
  {author} {\bibfnamefont {J.}~\bibnamefont {Wang}}, \bibinfo {author}
  {\bibfnamefont {Q.}~\bibnamefont {Chen}}, \bibinfo {author} {\bibfnamefont
  {J.}~\bibnamefont {Sun}}, \bibinfo {author} {\bibfnamefont {C.}~\bibnamefont
  {Li}}, \bibinfo {author} {\bibfnamefont {X.}~\bibnamefont {Wang}},\ and\
  \bibinfo {author} {\bibfnamefont {N.}~\bibnamefont {Gu}},\ }\href@noop {}
  {\bibfield  {journal} {\bibinfo  {journal} {Nanotechnology}\ }\textbf
  {\bibinfo {volume} {26}},\ \bibinfo {pages} {355501} (\bibinfo {year}
  {2015})}\BibitemShut {NoStop}%
\end{thebibliography}%
\end{document}